\documentclass[twocolumn,prb,showpacs]{revtex4-1}
\usepackage{graphicx}
\usepackage{dcolumn}
\usepackage{bm}
\usepackage{color}

\begin{document}
\title{Spin gap of the three-leg $S=3/2$ Heisenberg tube}
\author{S.~Nishimoto}
\affiliation{Institut f\"ur Theoretische Festk\"orperphysik, IFW Dresden, D-01171 Dresden, Germany}
\author{Y.~Fuji and Y.~Ohta}
\affiliation{Department of Physics, Chiba University, Chiba 263-8522, Japan}
\date{\today}
\begin{abstract}
The ground-state properties of the three-leg $S=3/2$ Heisenberg tube are studied using 
the density-matrix renormalization group method. We find that the spin-excitation gap 
associated with a spontaneous dimerization opens for the whole coupling regime, 
as seen in the three-leg $S=1/2$ Heisenberg tube. However, in contrast to the case of 
$S=\frac{1}{2}$ tube, the gap increases very slowly with increasing the rung coupling 
and its size is only a few $\%$ or less of the leg exchange interaction in the weak- and 
intermediate-coupling regimes. We thus argue that, unless the rung coupling is substantially 
larger than the leg coupling, the gap may be quite hard to be observed experimentally.
We also calcuate the quantized Berry phase to show that there exist three kinds of 
valence-bond-solid states depending on the ratio of leg and rung couplings.
\end{abstract} 
\pacs{75.10.Jm, 75.30.Kz, 75.40.Cx, 75.40.Mg}
\maketitle

For a long time, the exotic phenomena emerged from geometrical frustration have been 
fascinating but challenging subjects of research in condensed matter physics.~\cite{Moessner06} 
The peculiar dilemma of frustrated systems generally comes from a highly-degenerate 
ground state in the classical sense. Here, we know that to resolve it comes 
essentially back to how the degeneracy is removed or how the frustration is minimized 
by taking the quantum fluctuations into account. The simplest example for the spin frustration 
may be the 120$^\circ$ structure of antiferromagnetic triangle. In the context of triangular-lattice 
$S=1/2$ antiferromagnet a spin-liquid state was proposed by Anderson.~\cite{Anderson73} 
As a related issue, odd-leg spin tube systems such as Na$_2$V$_3$O$_7$ (Ref.~\onlinecite{Millet99}) 
and [(CuCl$_2$ tachH)$_3$Cl]Cl$_2$ (Ref.~\onlinecite{Seeber04}) have attracted much attention 
for the last few years. One could say that odd-leg spin ladders belong to the same universality 
class as single chain does; thus, the ground state is comprehended as a gapless 
spin liquid or a Tomonaga-Luttinger (TL) liquid.~\cite{Dagotto96} However, if the periodic boundary 
conditions (PBC) are applied in the rung direction, i.e., a tube is shaped, the spin states 
are dramatically changed due to spin frustration in the polygonal ring with odd number of rungs.

Quite recently, the hexagonal compound CsCrF$_4$ (Ref.~\onlinecite{Babel78}), which is 
an ideal three-leg spin tube system formed by Cr$^{3+}$ ions, has been reexamined experimentally 
from the point of view of spin frustration.~\cite{Manaka09} Since the magnetic moment comes 
from the {\it e}$_g^2$ state of Cr$^{3+}$ ion, the magnitude of spin on each site is $S=\frac{3}{2}$. 
By performing magnetic susceptibility, heat capacity $C(T)$, and electron spin resonance 
measurements, it was reported that no magnetic long-range order is observed down to $T=1.3$K. 
In particular, a gapless spin-liquid state (or a TL liquid state) is indicated from 
the finite $T$-linear component of $C(T)$; this result raises a more absorbing question 
because a gapped ground state is expected in odd-leg spin-half-integer spin tube system.~\cite{Schulz96} 
The need for an investigation of odd-leg spin tube system with $S=\frac{3}{2}$ 
is therefore obvious in order to figure out this puzzle. 

\begin{figure}[t]
    \includegraphics[width= 5.0cm,clip]{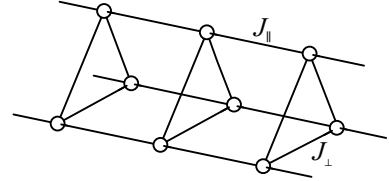}
  \caption{
Lattice structure of three-leg Heisenberg tube.
  }
    \label{lattice}
\end{figure}

In this paper, we thus consider three-leg $S=\frac{3}{2}$ Heisenberg tube. The Hamiltonian is given as
\begin{equation}
H = J_\parallel \sum_{\alpha=1}^3 \sum_i \vec{S}_{\alpha,i} \cdot \vec{S}_{\alpha,i+1} 
+ J_\perp \sum_i \sum_{\alpha (\neq \alpha^\prime)} \vec{S}_{\alpha,i} \cdot \vec{S}_{\alpha^\prime,i},
\label{hamiltonian}
\end{equation}
where $\vec{S}_{\alpha,i}$ is a spin-$\frac{3}{2}$ operator at leg $\alpha(=1,2,3)$ and rung $i$. 
$J_\parallel$ and $J_\perp$ are the nearest-neighbor exchange interactions in the leg and rung directions, 
respectively (see Fig.\ref{lattice}). We take $J_\parallel=1$ as the unit of energy hereafter.
In order to investigate the ground-state properties of the system (\ref{hamiltonian}), 
the density-matrix renormalization group (DMRG) technique~\cite{White92} is employed. 
As necessary, the PBC or the open-end boundary conditions (OBC) are chosen in the leg direction. 
Using the OBC (PBC), we study the tubes with several kinds of length $L = 12$ to $48$ ($L = 8$ to $24$) 
keeping $m = 1200$ to $2600$ ($m = 1600$ to $3200$) density-matrix eigenstates in the 
renormalization procedure; in this way, the typical truncation error, i.e., the discarded
weight, is $2 \times 10^{-6}-1 \times 10^{-5}$ ($3 \times 10^{-5}-7 \times 10^{-5}$). 
We note that the system length must be even and is better to be kept in $L=4l$ or $4l+2$ 
with $l$=integer for systematic extrapolation of calculated quantities into the thermodynamic limit. 
Moreover, an extrapolation to $m \to \infty$ for each calculation is necessary because the DMRG 
trial wave function slowly approaches the exact one with increasing $m$ due to the large degrees 
of freedom $\sim 4^{3L}$ and strong spin frustration in our system. All the calculated quantities 
in this paper are extrapolated to the limit $m \to \infty$; thus, for example, the maximum error 
in the ground-state energy is estimated to be less than $1 \times 10^{-3}$.

\begin{figure}[t]
    \includegraphics[width= 6.2cm,clip]{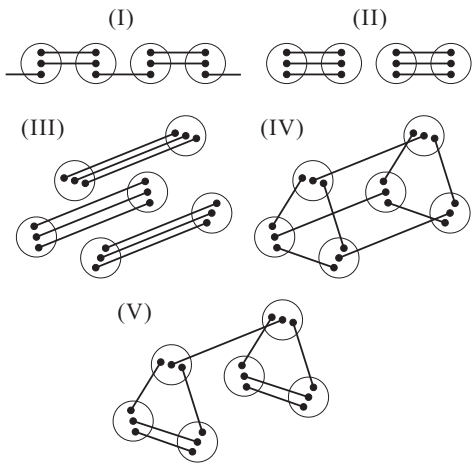}
  \caption{
Schematic pictures of the VBS configurations. Each small solid circle and connection 
with line denote a spin-$\frac{1}{2}$ variable and a singlet pair, respectively. The large open 
circle represents a spin-$\frac{3}{2}$ operation which symmetrizes three spin-$\frac{1}{2}$ 
variables. The configurations (I) and (II) are VBS states in small and large 
bond-alternation regimes of the 1D $S=\frac{3}{2}$ Heisenberg model, respectively. 
The configurations (III)-(V) are possible candidates of VBS states in the three-leg 
$S=\frac{3}{2}$ Heisenberg tube (details are given in the text and Fig.~\ref{PD}).
   }
    \label{RVB}
\end{figure}

The first thing we think of when considering the system (\ref{hamiltonian}) might be 
the topological similarity to the other odd-leg half-integer-spin Heisenberg tubes. 
The simplest case, i.e., three-leg $S=\frac{1}{2}$ Heisenberg tube, has been well-studied, 
and the ground state is known to be gapped where the system is spontaneously dimerized 
in the leg direction to relax the intra-rung spin frustration.~\cite{Schulz96,Kawano97,Nishimoto08,Sakai08}
This can be naturally understood by analogy with the gap-opening mechanism 
in the one-dimensional (1D) $S=\frac{1}{2}$ spin-Peierls Heisenberg model.~\cite{Bray83} Hence, 
in the case of $S=\frac{3}{2}$ as well, it would be best to start with bond-alternated 
single chain problem, namely, the 1D $S=\frac{3}{2}$ spin-Peierls Heisenberg model. 
The Hamiltonian is written as $H=\sum_{i=1}^{L-1}[1-(-1)^i\delta]\vec{S}_i \cdot \vec{S}_{i+1}$ 
where $\vec{S}_i$ is a spin-$\frac{3}{2}$ operator at site $i$ and $\delta (>0)$ is the strength 
of bond alternation. The low-energy physics of this system has been fundamentally 
elucidated:~\cite{Yajima96,Yamamoto97} Across the critical point $\delta \approx 0.42$, 
two kinds of valence-bond-solid (VBS) phases appear in the ground state; 
for the larger alternation ($\delta>0.42$), the VBS state is essentially written as a direct product 
of `simple' spin-Peierls singlet bonds [Fig.~\ref{RVB}(II)], whereas 
for the smaller alternation ($\delta<0.42$), it is expressed as a combined state of the spin-Peierls 
singlet and $S=1$ Haldane-like-gapped configurations [Fig.~\ref{RVB}(I)]. And, the ground state is 
always gapped except at the critical point. Now therefore, getting back to our system (\ref{hamiltonian}), 
if the spontaneous dimerization occurs as in the three-leg $S=\frac{1}{2}$ Heisenberg tube, 
it is likely that a gapped ground state is also obtained here.

\begin{figure}[t]
    \includegraphics[width= 6.2cm,clip]{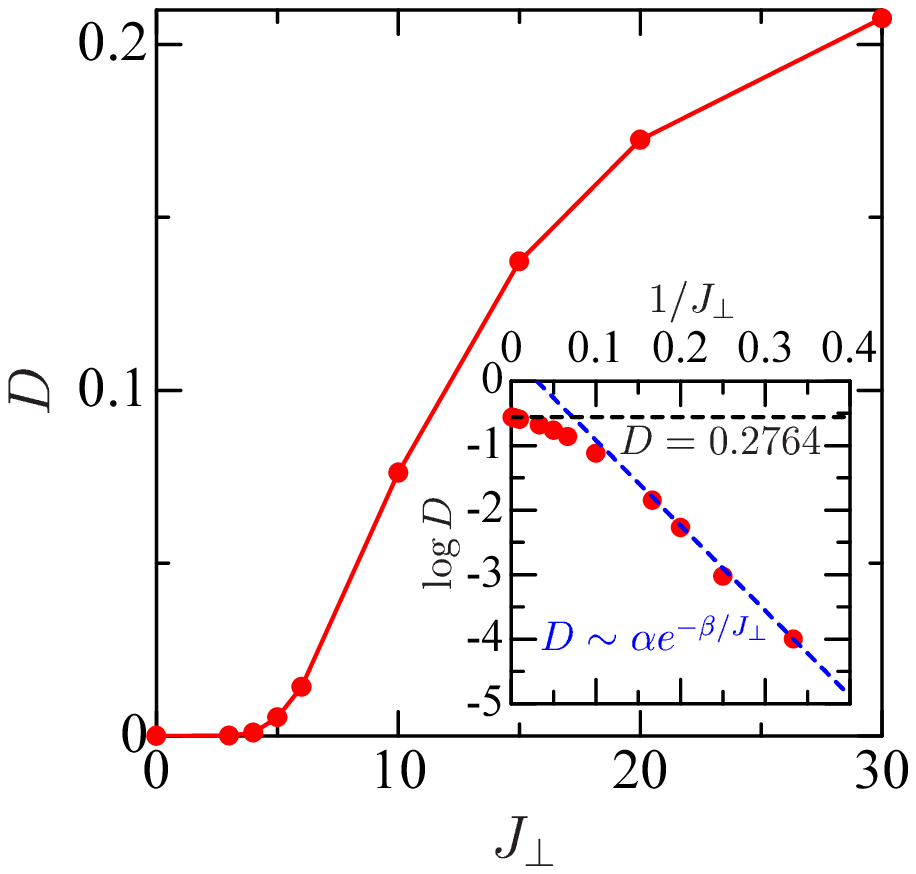}
  \caption{
Dimerization order parameter $D$ as a function of $J_\perp$. Inset: Semilog plot of $D$ vs. 
$1/J_\perp$. $D$ saturates to $0.2764$ in the large $J_\perp$ limit. The data for small $J_\perp$ is 
fitted by a function $D=\alpha \exp (-\beta/J_\perp)$ with $\alpha=0.25$ and $\beta=13.2$. 
}
    \label{dimerOP}
\end{figure}

Then, we will simply evaluate a dimerization order parameter to check the presence or absence 
of long-ranged spin-Peierls ordering in our system. Since a spin-Peierls state is characterized 
as an alternating formation of spin-singlet pairs in the leg direction, we focus on the nearest-neighbor 
spin-spin correlations,
\begin{equation}
S(i) = -\left\langle \vec{S}_{\alpha,i} \cdot \vec{S}_{\alpha,i+1} \right\rangle,
\end{equation}
where $\left< \cdots \right>$ denotes an expectation value in the ground state. Note that 
this quantity is independent of $\alpha$. With applying the OBC, the translational symmetry is 
broken due to the Friedel oscillation and the spin-Peierls state is directly observable as 
a ground state. In general, the Friedel oscillation in the center of the system decays as 
a function of the system length. If the amplitude at the center of the system
\begin{equation}
D(L) = \left|S(L/2) - S(L/2+1)\right|
\end{equation}
persists for arbitrarily long system length, it corresponds to a long-ranged dimerization 
order which indicates the spin-Peierls ground state. The order parameter is 
thus defined as an extrapolated value into the thermodynamic limit,
\begin{equation}
D=\lim_{L \to \infty} D(L).
\end{equation}

In Fig.~\ref{dimerOP}, the extrapolated values $D$ are plotted as a function of $J_\perp$. 
We see that it increases gradually at $J_\perp \lesssim 5$, almost linearly at $5 \lesssim J_\perp \lesssim 15$, 
and then go into saturation at $D = 0.2764$ in the large $J_\perp$ limit. These different 
behaviors could be interpreted in terms of different VBS state for each the $J_\perp$ regime, 
as in the $S=\frac{1}{2}$ Heisenberg tube.~\cite{Nishimoto08} This point will be clarified 
below by examining the Berry phase. A remaining question would be whether 
the order parameter remains finite for small $J_\perp$ regime (we have not successfully 
obtained $D$ for $J_\perp<3$ due to its smallness). We find that the behavior seems just 
like the Berezinskii-Kosterlitz-Thouless type transition, $D = 0.25 \exp(-13.2/J_\perp)$, 
from the single logarithmic plot (see the inset of Fig.~\ref{dimerOP}); it may imply that 
the order parameter is exponentially small but finite at $0<J_\perp<3$. 
We conclude that in the wide range of $J_\perp$ the dimerization order occurs and 
the spin-excitation gap is expected to be finite there.

\begin{figure}[t]
    \includegraphics[width= 6.2cm,clip]{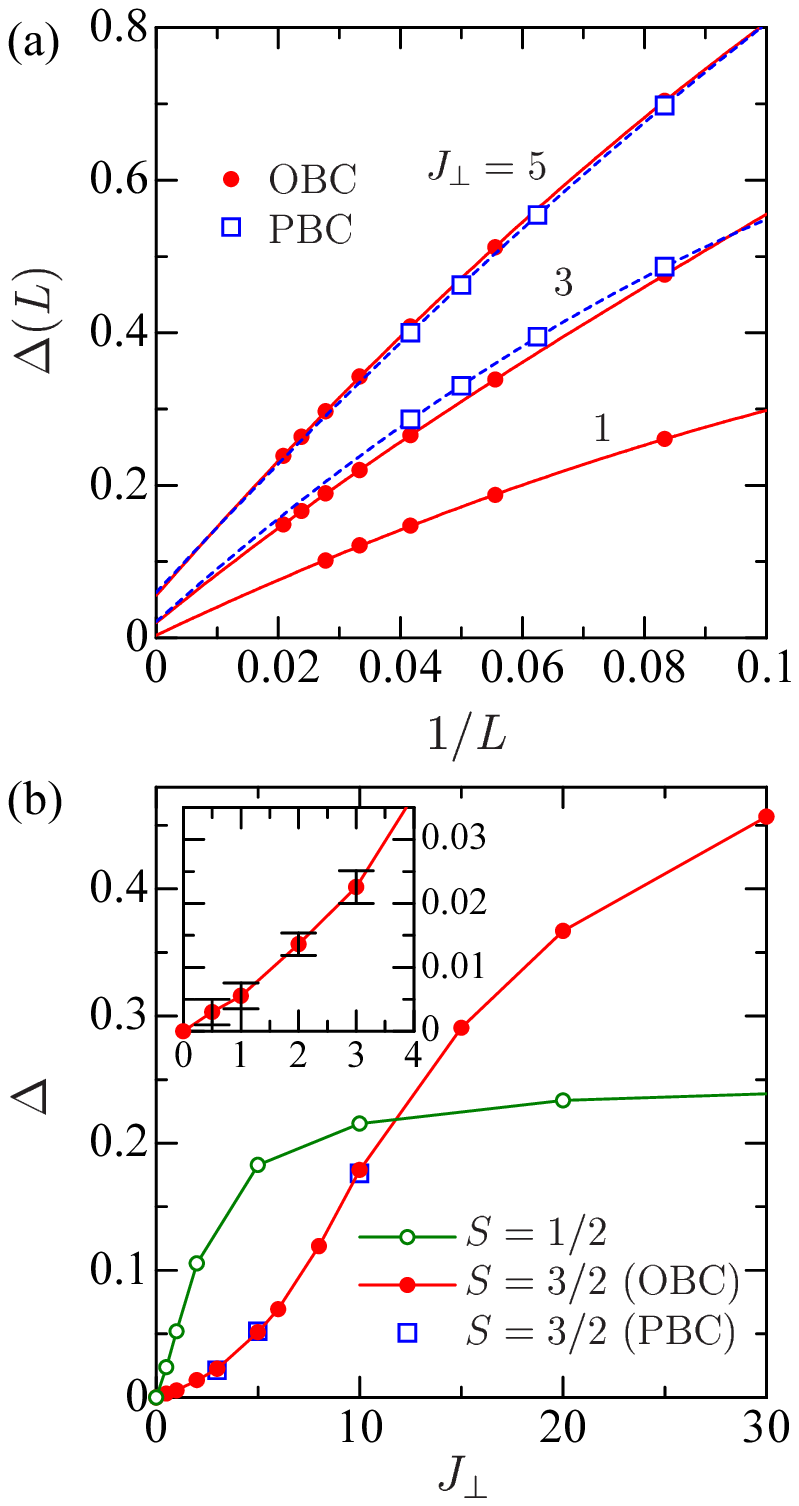}
  \caption{
(a) Finite-size scaling of $\Delta(L)$ as a function of $1/L$. The lines are the polynomial 
fittings. (b) Extrapolated values of $\Delta(L)$ to the thermodynamic limit $1/L \to 0$. 
Inset: extended figure for $0 \le J_\perp \le 4$. The error bars give differences between 
the second-order and cubic polynomial fittings for $\Delta(L)$.
}
    \label{gap}
\end{figure}

Let us then estimate the spin-excitation gap. Of particular interest is the evolution of 
the gap onto the ratio between leg and rung couplings. The gap is defined as
\begin{equation}
\Delta=E_1(L)-E_0(L), \ \ \ \Delta=\lim_{L \to \infty}\Delta(L),
\end{equation}
where $E_0(L)$ and $E_1(L)$ are energies of the ground state ($S=0$) and first triplet 
excited state ($S=1$) for the system with length $L$, respectively. 
In Fig.~\ref{gap}(a), we plot the system-size dependence of the spin-excitation gap 
calculated with the OBC in full circles. We see that the values of $\Delta(L)$ can be 
smoothly extrapolated to the thermodynamic limit $1/L \to \infty$. 
The extrapolated values $\Delta$, using a cubic polynomial extrapolation 
for $\Delta(L)$, are shown in Fig.~\ref{gap}(b) as a function of $J_\perp$. 
As expected, $J_\perp$-dependence of $\Delta$ looks rather similar to that of $D$; namely, 
it increases slowly at $J_\perp \lesssim 5$, rapidly at $5 \lesssim J_\perp \lesssim 15$, 
and then saturates to $\Delta = 0.6661$ in the strong-coupling limit $J_\perp=\infty$. 
This is because the spin-excitation gap is essentially equivalent to a binding energy of 
most weakly bounded spin-singlet pair in the system and it is approximately scaled with 
the dimerization strength. 

Here, it is very instructive to compare the gap with that of the $S=\frac{1}{2}$ tube, 
which is also shown in Fig.~\ref{gap}(b). Two remarks are made by the comparison: 
(i) Although it may be rather natural, the gap in the strong-coupling limit seems to be scaled 
with the magnitude of spin, $\Delta(J_\perp \to \infty) \propto S$, (ii) the gap for $S=\frac{3}{2}$ 
increases much more slowly with increasing $J_\perp$ in the weak-coupling regime. 
As a result, only a few $\%$ of the leg exchange interaction remains even at $J_\perp/J_\parallel=5$. 
Hence, we argue that, unless the ratio $J_\perp/J_\parallel$ is sizably large, it may 
be difficult to detect the gap experimentally. For CsCrF$_4$, the leg exchange interaction 
is estimated to be a few 10 to 100 K by comparing the experimental peak position of 
magnetic susceptibility to numerical analysis for the 1D $S=\frac{3}{2}$ Heisenberg chain,~\cite{Jongh74} 
and the gap is only a few K at the outside even for $J_\perp/J_\parallel=5$. 

As described above, we obtain the gapped ground state for the whole $J_\perp$ region 
with applying the OBC. It would mean that our system never contain the Haldane-type 
VBS state [Fig.~\ref{RVB} (I)] in the three chains. This is because the gap cannot open 
due to free edge spins created with the OBC if the state (I) is included. But to be sure, 
we shall confirm it by estimating the gap with the PBC. The obtained results are shown 
with open squares in Fig.\ref{gap}. We see that the extrapolated values are in good agreement 
to those with the OBC and it is confirmed that the state (I) does not exist at any VBS state 
in our system. Now, it is a fair question then to ask which kind of VBS state is formed 
and how it changes with varying $J_\perp$.

\begin{figure}[t]
    \includegraphics[width= 6.5cm,clip]{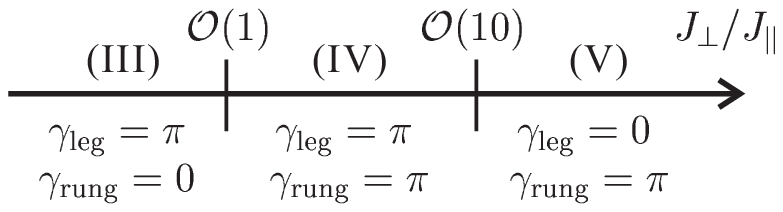}
  \caption{
Schematic phase diagram of the three-leg $S=\frac{3}{2}$ Heisenberg tube, classified 
by the Berry phases on the leg bond ($\gamma_{\rm leg}$) and rung bond ($\gamma_{\rm leg}$). 
The Roman numbers correspond to the VBS states shown in Fig.~\ref{RVB}.
}
    \label{PD}
\end{figure}

Finally, we investigate the quantized Berry phase for determining topological configuration 
of VBS ground state. The Berry phase is defined by
\begin{equation}
\gamma=-i\int^{2\pi}_{0}A(\phi)d\phi,
\end{equation}
where $A(\phi)$ is the Abelian Berry connection, 
$A(\phi)=\langle \psi_\phi | \partial_\phi \psi_\phi \rangle$ 
with the ground state $|\psi_\phi \rangle$.~\cite{Hatsugai06} The Berry phase is quantized 
as $0$ or $\pi$ (mod $2\pi$) if the system has spin gap during the adiabatic continuation 
and time reversal symmetry; and ``undefined'' if a gapless excitation exists. 
We introduce a local perturbation by a local twist of the nearest-neighbor connection,
$
\vec{S}_{\alpha,i} \cdot \vec{S}_{\alpha^\prime,j} \to 
\frac{1}{2}(e^{-i\phi}S_{\alpha,i}^+S_{\alpha^\prime,j}^-
+e^{i\phi}S_{\alpha,i}^-S_{\alpha^\prime,j}^+)+S_{\alpha,i}^zS_{\alpha^\prime,j}^z.
$
In this paper, a couple of clusters with $L=2$ and $4$ are analyzed for this quantity. 
A dimerized pair of triangles, including six spins, from the clusters are picked up (it is 
the cluster itself for $L=2$), and the Berry phases of the leg bond ($\gamma_{\rm leg}$) 
for $\alpha=\alpha^\prime, j=i+1$ and of the rung bond ($\gamma_{\rm rung}$) 
for $\alpha \neq \alpha^\prime, j=i$ are evaluated. We call the spin-singlet pair on 
the leg (rung) bond ``on-leg (on-rung) pair''.

The Berry phases are obtained as follows: $(\gamma_{\rm leg},\gamma_{\rm rung})=(\pi,0)$ 
at $0<J_\perp<1$ ($0<J_\perp<0.5$), $(\gamma_{\rm leg},\gamma_{\rm rung})=(\pi,\pi)$ at 
$1<J_\perp<15.3$ ($1<J_\perp<18$), and $(\gamma_{\rm leg},\gamma_{\rm rung})=(0,\pi)$ 
at $J_\perp>15.3$ ($J_\perp>18$) for $L=2$ ($L=4$) cluster. Accordingly, we find three different 
phases depending on $J_\perp/J_\parallel$, as shown in Fig.~\ref{PD}. The term ``phase transition'' 
describes a recombination of VBS bonds. In the large-coupling regime $J_\perp/J_\parallel > {\cal O}(10)$, 
we can easily imagine that the on-rung pair is more stable than the on-leg pair 
and as many pairs as possible prefer to be formed on the rung bond [Fig.~\ref{RVB}(V)]. 
The spin gap is therefore scaled with the binding energy of on-leg pair, i.e., $\Delta \propto J_\parallel$, 
which is consistent with the saturating behavior of $\Delta$ for $J_\perp \gg J_\parallel$. 
On the other hand, in the weak-coupling regime $J_\perp/J_\parallel < {\cal O}(1)$, 
all spin-singlet pairs are formed on the leg bond [Fig.~\ref{RVB}(III)] because the binding energy 
of the on-leg pair is much larger than that of the on-rung pair. And, in the intermediate regime 
${\cal O}(1) < J_\perp/J_\parallel < {\cal O}(10)$, the spin-singlet pairs seem to be 
distributed {\it in a balanced manner} [Fig.~\ref{RVB}(IV)]. Here, we notice an interesting 
relation to the phase transition in the $S=\frac{1}{2}$ tube.~\cite{Nishimoto10} If we ignore 
a spin-singlet pair on each rung in the phases (IV) and (V) of our system, the remaining 
degrees of freedom are equivalent to those of the $S=\frac{1}{2}$ tube. As it turns out, 
the recombination of the remaining VBS bonds between (IV) and (V) can be essentially equivalent 
to the phase transition in the $S=\frac{1}{2}$ tube. Then the (ignored) extra degrees of freedom 
yields the additional phase (III) in our $S=\frac{3}{2}$ system.

In conclusion, we study the ground-state properties of the three-leg $S=3/2$ Heisenberg tube 
with the DMRG method. It is confirmed that a spontaneous dimerization occurs and 
the spin-excitation gap opens for the whole coupling region. This may be a common feature of 
odd-leg half-integer-spin Heisenberg tube systems. 
We find that the gap for $S=\frac{3}{2}$ increases very slowly with increasing $J_\perp$ 
and it remains very small compared with $J_\parallel$ in the weak- to intermediate-coupling regions. 
For CsCrF$_4$, the gap is estimated to be only a few K or less at normal pressures and, 
for example, additional condition such as applying pressure might be required to enlarge the ratio 
$J_\perp/J_\parallel$ in order to detect the gapped state. 
Moreover, by calculating the quantized Berry phase, it is shown that two phase transitions 
as recombination of VBS bonds occur with varying the ratio $J_\perp/J_\parallel$ 
although further work is desirable for quantitative evaluation of the critical points of 
the phase transitions.

\acknowledgments

This work was supported in part by Kakenhi Grant No.~22540363 of Japan. 
A part of computations was carried out at Research Center for Computational 
Science, Okazaki.

\end{document}